\documentclass[12pt]{IEEEtran}
\pdfoutput=1

\usepackage[version=3]{mhchem}

\usepackage{color}

\usepackage{amsmath,amssymb,amsfonts}
\usepackage{algorithmic}
\usepackage{graphicx}
\usepackage{textcomp}
\usepackage[binary-units=true,
    prefixes-as-symbols=true, separate-uncertainty = true,multi-part-units=single]{siunitx}
\DeclareSIUnit\year{yr}

\usepackage[percent]{overpic}
\usepackage{transparent}

\usepackage{cleveref}
\renewcommand{\cref}{\Cref}
\newcommand{\eg}{e.g.}
\newcommand{\ie}{\emph{i.e.}}

\newcommand{\compair}{\emph{ComPair}}
\newcommand{\amego}{\emph{AMEGO}}

\newcommand{\eplus}{$e^+$}
\newcommand{\eminus}{$e^-$}

\usepackage{wrapfig}
\usepackage[export]{adjustbox}
\begin{document}

\title{Development of a Silicon Tracker for the\\ All-sky Medium Energy Gamma-ray Observatory Prototype}

\author{Sean~Griffin$^1$\thanks{$^1$UMCP / NASA GSFC ; \href{mailto:sean.griffin@nasa.gov}{sean.griffin@nasa.gov}} for the AMEGO Team$^2$\thanks{$^2$\href{https://asd.gsfc.nasa.gov/amego/}{https://asd.gsfc.nasa.gov/amego/}}}

\IEEEtitleabstractindextext{%
\begin{abstract}
    The gamma-ray sky from several hundred keV to $\sim$ a hundred MeV has remained largely unexplored due to the challenging nature of detecting gamma rays in this regime. 
    At lower energies, Compton scattering is the dominant interaction process whereas at higher energies pair production dominates, with a crossover at a few MeV. 
    Thus, an instrument designed to work in this energy range must be optimized for both Compton and pair-production events.
    \amego, the \emph{All-sky Medium Energy Gamma-ray Observatory}, a Probe-class mission in consideration for the 2020 decadal survey, is designed to operate at energies from $\sim\SI{200}{\keV}$ to $>$ \SI{10}{\GeV} with over an order of magnitude increase in sensitivity and with superior angular and energy resolution compared to previous instruments. 
    \amego\ comprises four major subsystems: a plastic anti-coincidence detector for rejecting cosmic-ray events, a silicon tracker for tracking pair-production products and tracking and measuring the energies of Compton-scattered electrons, a cadmium-zinc-telluride (CZT) calorimeter for measuring the energy and location of Compton scattered photons, and a CsI calorimeter for measuring the energy of the pair-production products at high energies. 
    A prototype instrument, known as \emph{ComPair}, is under development at NASA's Goddard Space Flight Center and the US Naval Research Laboratory.
    In this contribution, we provide details on the development of the silicon tracker subsystem. 
\end{abstract}

}


\maketitle
\IEEEpeerreviewmaketitle
\IEEEdisplaynontitleabstractindextext

\section{Context}

\begin{figure}[hbtp]
\centering
\includegraphics[width=0.50\textwidth]{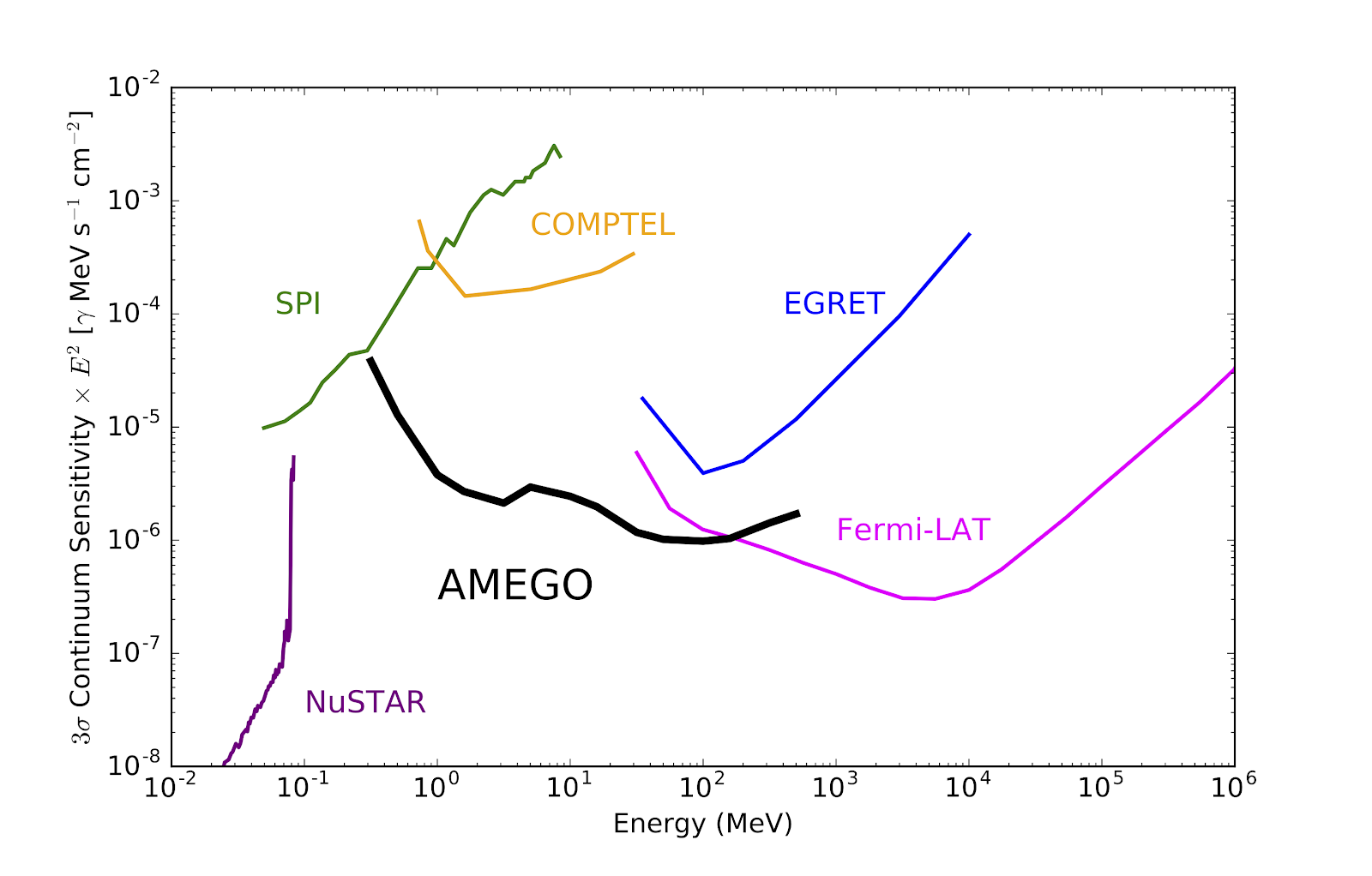}
\caption{\amego\ will provide over an order-of-magnitude increase in sensitivity over other instruments in the \SI{}{\MeV} band \cite{amegosims}.
}
\label{fig:AMEGOsensitivity}
\end{figure}

To date, the region of the electromagnetic spectrum between \SI{100}{\keV} and \SI{100}{\MeV} has had very limited astrophysical observations. 
One of the difficulties in constructing an instrument to operate at these energies is that around a few hundred \SI{}{\keV}, gamma-ray interactions are dominated by Compton scattering, and at higher energies \eplus / \eminus\ pair production is the dominant process, with a crossover at a few \SI{}{\MeV}.
An instrument designed to work in this energy regime must thus be optimized for both Compton and pair-production events; several designs have been proposed in the past to accomplish this (e.g. \cite{MEGA, grips}). 
\IEEEpubidadjcol 
\amego\ is a NASA-led Probe-class mission concept being submitted to the 2020 Decadal Survey. 
\amego\ will provide an order-of-magnitude increase in sensitivity above previous measurements made in the \SI{}{\MeV} band (see \cref{fig:AMEGOsensitivity}), will be sensitive to polarization, and will have sufficiently high energy resolution for spectroscopic measurements of nuclear emission lines. 
An overview of the instrument design and simulated performance can be found in Refs. \cite{amegoprobe} and \cite{amegosims}, respectively.

\begin{figure}[hbtp]
\centering
\includegraphics[width=0.50\textwidth]{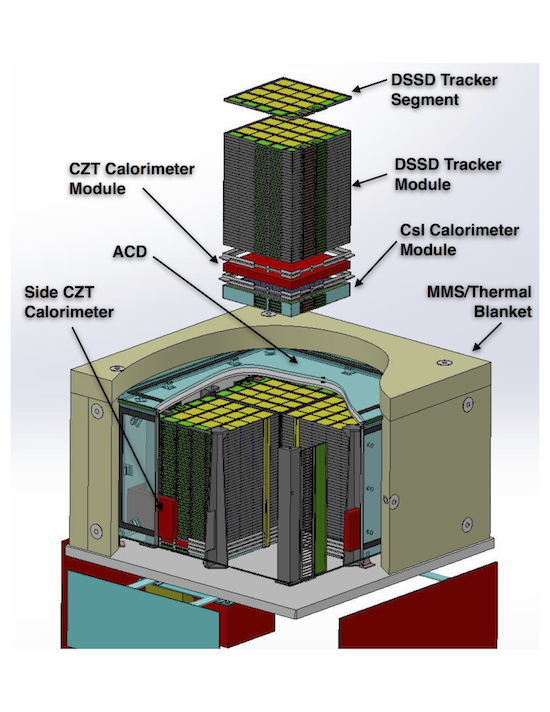}
\vspace{-1cm}
\caption{Hardware design for \amego\ \cite{amegoprobe}.
}
\label{fig:AMEGOdesign}
\end{figure}

A schematic of the hardware design is given in \cref{fig:AMEGOdesign}.
\amego\ comprises four major subsystems: a plastic anticoincidence detector for vetoing cosmic-ray events, a silicon tracker system for measuring particle tracks and energy deposition (the subject of this work), a cadmium-zinc-telluride (CZT) calorimeter for measuring the direction and energy of Compton scattered photons \cite{cztcal}, and a CsI calorimeter for measuring the position and energy of pair-production events \cite{csical}. 

\section{The Silicon Tracker Subsystem}
\subsection{Design}

The \amego\ tracker comprises 60 layers of double-sided silicon strip detectors (DSSDs) vertically separated by \SI{1}{\centi\meter} connected in a $4\times 4$ configuration, referred to as a ``tower''.
In the current design, each detector is a \SI{9.5}{\centi\meter} square, \SI{500}{\micro\meter} thick with a strip pitch of \SI{500}{\micro\meter}, yielding 192 strips per side.
These parameters, in addition to the vertical spacing between layers (\SI{1}{\cm}) dictate the angular resolution, field of view, and collection area of the telescope.
The detector readout electronics are located along two edges of each segment (one for each the $x$ and $y$ directions) with a minimal amount of passive material between detector elements. 
The full \amego\ tracker has four identical towers; this modular design simplifies integration and allows for the easy production of spare components.
Furthermore, this helps minimize the amount of passive material within the instrument, which is crucial to operation in the Compton regime. 
ASICs are used to read out the strips; the ASICs provide pulse-height information which is required to measure the energy of the Compton recoil electron, for measuring the energy of the particles in low-energy pair-production events, and for identifying pair-production particle tracks for event reconstruction.

A prototype instrument, known as \compair\ (Compton/Pair-Production), is currently being developed at NASA Goddard and the Naval Research Laboratory. 
The prototype will include small versions of each of the four \amego\ subsystems and has an ultimate goal of flying as a balloon payload. 
When fully assembled, the prototype tracker will comprise 6-10 single-detector layers; the development of this subsystem is discussed in the following sections.

\subsection{Detector Development}

A number of DSSDs have been purchased from Micron Semiconductor \cite{micron} and are currently being fabricated; several test units have arrived at Goddard and are currently undergoing characterization and testing. 
Each detector is a $\SI{10}{\cm}$ square and \SI{500}{\micro\meter} thick, with 192 strips per side, each with a pitch and width of $\sim \SI{510}{\micro\meter}$ and \SI{60}{\micro\meter}, respectively. 

The first few silicon detectors will be thoroughly characterized and tested: bulk current-versus-voltage (IV) and capacitance-vs-voltage (CV) measurements will be made of each detector, and single strip measurements of components such as the bias resistors and coupling capacitors will be made and compared against measurements provided by Micron.
This testing protocol is expected to be very time consuming (up to two weeks per detector); thus results from the first few wafers will define the testing strategy for subsequent wafers. 
It is likely that subsequent silicon wafers will be ``bulk tested''; by combining signals of $N$ strips while testing, the testing time is reduced by a factor of $\sim N$. 
Thus, every strip is still tested but the process is far less time consuming; this is important since when fully assembled, \amego\ would have some 370 thousand strips, making it impractical to test every single one. 
Using this method, the general vicinity of bad strips can be identified by comparing signals from bulk-tested strips with expected signals based on the single-channel results made on another detector. 

\begin{figure}[hbtp]
\centering
\includegraphics[width=0.48\textwidth]{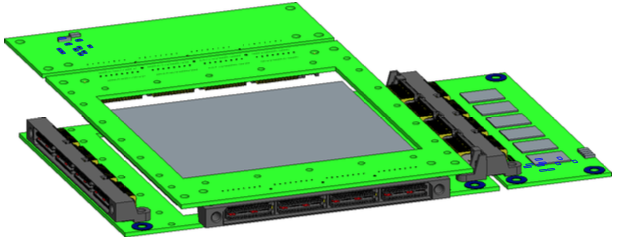}
\includegraphics[width=0.48\textwidth]{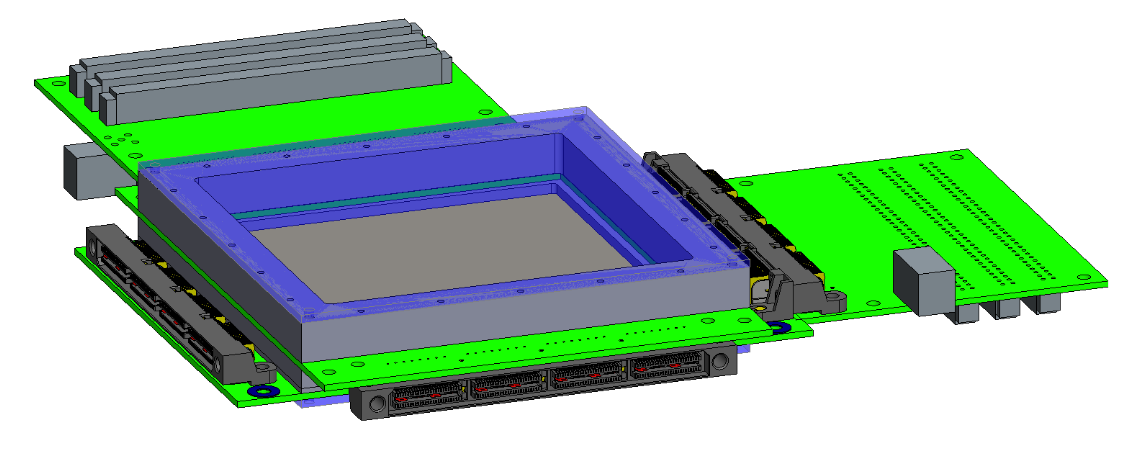}
\includegraphics[width=0.48\textwidth]{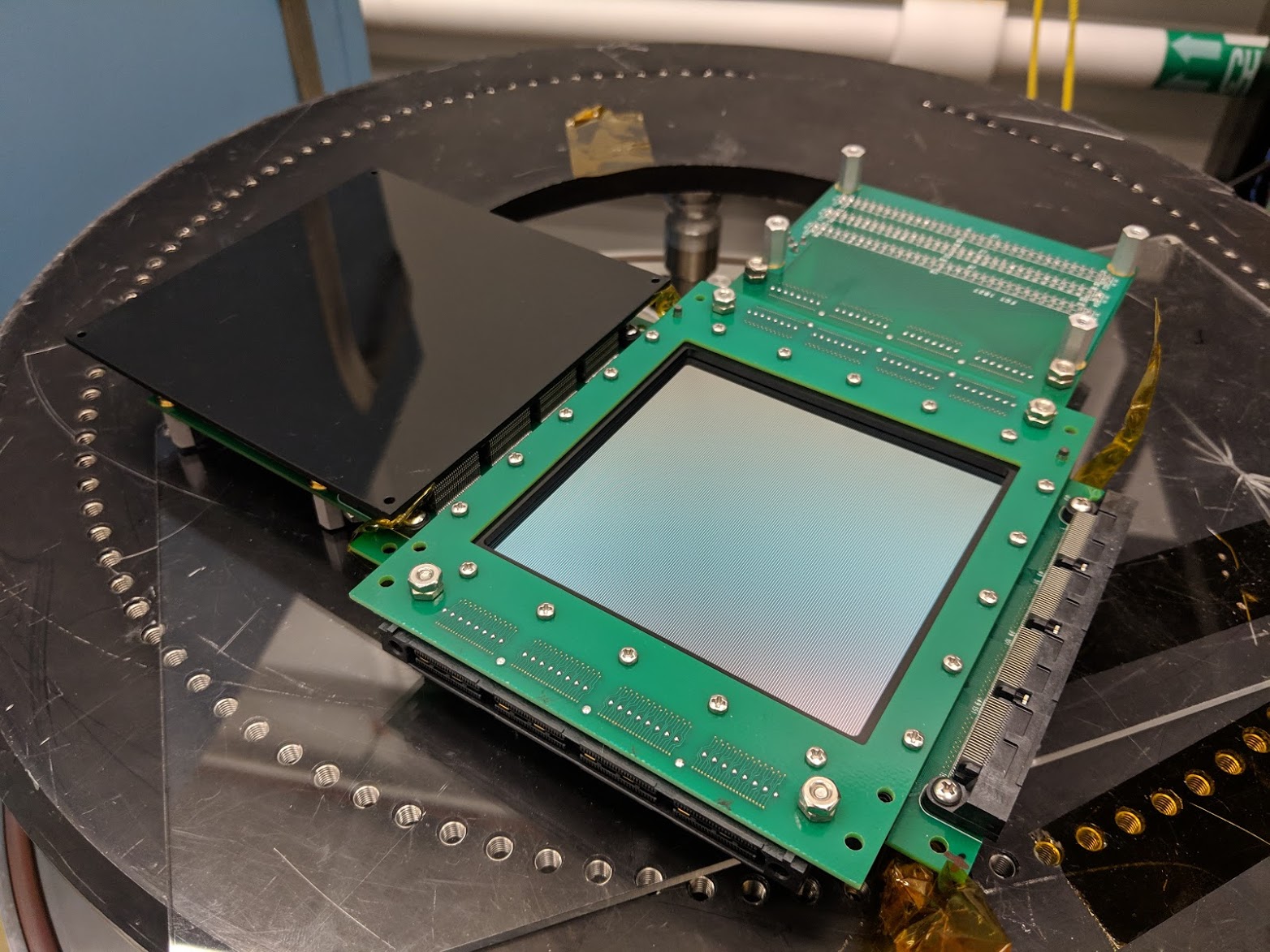}
\caption{\textbf{Top:} CAD model of the single-wafer carrier board printed wire boards (PWBs) with a mock-up of an ASIC readout board (right side, populated with six ASICs) mated to one of the connectors. 
Each carrier is made up of two of these boards with one flipped and rotated $90^\circ$. 
The silicon detector (gray) can be seen through the window in the carrier board. 
\textbf{Middle:} CAD model of the carrier including Delrin structural material; in this version a 192-pin breakout board is mated to the connectors.
\textbf{Bottom:} Photo of the carrier board populated with a DSSD. 
The black plate normally covers the DSSD to provide some amount of light protection and to prevent accidental damage to the detector. 
The bottom side of a signal breakout board can be seen mated to the top-left connector. 
}
\label{fig:carrier}
\end{figure}

Single-detector ``carriers'' have been designed to encapsulate and provide an interface to both sides of the detectors (required for both biasing and readout). 
A CAD image of the carrier design is given in \cref{fig:carrier} alongside a photo of a populated carrier. 
Each assembly uses two identical printed wire boards, with one flipped and rotated $90^\circ$ to read out the orthogonal strips. 
The signals from the DSSDs are captured using elastomer connectors between the carrier boards and the detector, with a plastic spacer and base plate to provide rigidity. 
This design does not require any wire-bonding and allows detectors to be removed from the carriers if need be.
Connectors are located on all four sides of the carrier and allow for multiple carriers to be connected (thus daisy-chaining the DSSDs) in both the $x$ and $y$ directions. 
The same connectors are used to interface with custom readout boards which contain ASICs and an FPGA for handling triggering (internal and external) and the readout of the individual ASICs. 

Finally, performance measurements of a daisy-chained DSSD ladder will be made to quantify the total parasitic capacitance and provide insight into the performance of a large DSSD array in preparations for the full \amego\ design. 
The instrument must minimize the total amount of passive material within it; this is particularly important in the Compton regime where measuring the energy of the scattered electron is critical to event reconstruction. 
Due to to total number of readout channels, using smaller detector arrays (\eg\ $2\times 2$ detector ladders) is also impractical. 
Thus, understanding the noise performance of a large DSSD ladder is crucial to the development of a future Compton/pair-production telescope.

When daisy-chained, the capacitance of each DSSD adds linearly and the readout noise scales linearly with load capacitance (this is demonstrated in the next section), so a ladder will have significantly higher noise than a single detector. 
These measurements will be accomplished by daisy-chaining detectors into an ``L-shape'' with four detectors to a side (requiring seven total), mimicking a full $4\times 4$ \amego\ tracker layer. 
The strips will be read out at the far end of each leg in the L-shape; position and energy resolution measurements of the corner element will be compared against that of a single detector.

\subsection{Front-End Development}

The ASIC currently under consideration for the \amego\ tracker is the IDEAS VATA460.3 \cite{ideas_asic}.
A development board containing one of these ASICs is undergoing testing at NASA Goddard to verify noise, power consumption, linearity, and maximum trigger rates. 
This ASIC is a highly configurable, 32-channel, combined preamplifier and 10-bit pulse-height-measuring digitizer, supporting both positive and negative polarity input pulses\footnote{Both polarities are required, since the polarity of the pulses from the DSSDs depends on whether the signals are from strips on the ohmic or junction side of the detector.}.
The maximum \emph{analog} dynamic range is $-\SI{90}{\femto\coulomb}$ ($+\SI{50}{\femto\coulomb}$) which corresponds to an energy deposition of $\sim\SI{2.0}{\MeV}$ ($\SI{1.1}{\MeV}$) in silicon\footnote{For positive charge, the maximum charge deposition can also be increased to \SI{90}{\femto\coulomb} albeit with reduced linearity.}. 
The \emph{digital} dynamic range (\ie\ the subset of the analog dynamic range covered by the digitizer) can be adjusted using configuration registers.
The ASIC has a global trigger threshold and each channel can be fine-tuned to account for channel-to-channel differences.

The noise performance of the tracker ASIC is critical to the performance of \compair\ and \amego, in particular at low energies where the energy of Compton scattered electrons is small. 
The noise in the readout electronics due to parasitic capacitance in the silicon detectors and associated interfaces scales linearly with load capacitance. 
It is possible to measure this behavior by loading an ASIC input with capacitors of different values and measuring the width of the fluctuations about the pedestal. 
For these measurements, a simple test board was made with selectable capacitor values with which to load an input channel, and the ASIC was externally triggered at \SI{500}{\hertz} for \SI{240}{\second} to provide sufficient statistics. 
The measured ADC values were binned; the mean of the distribution is the pedestal for that channel and the width (RMS) corresponds to the noise. 

\begin{figure}[hbtp]

\centering
\includegraphics[width=0.5\textwidth]{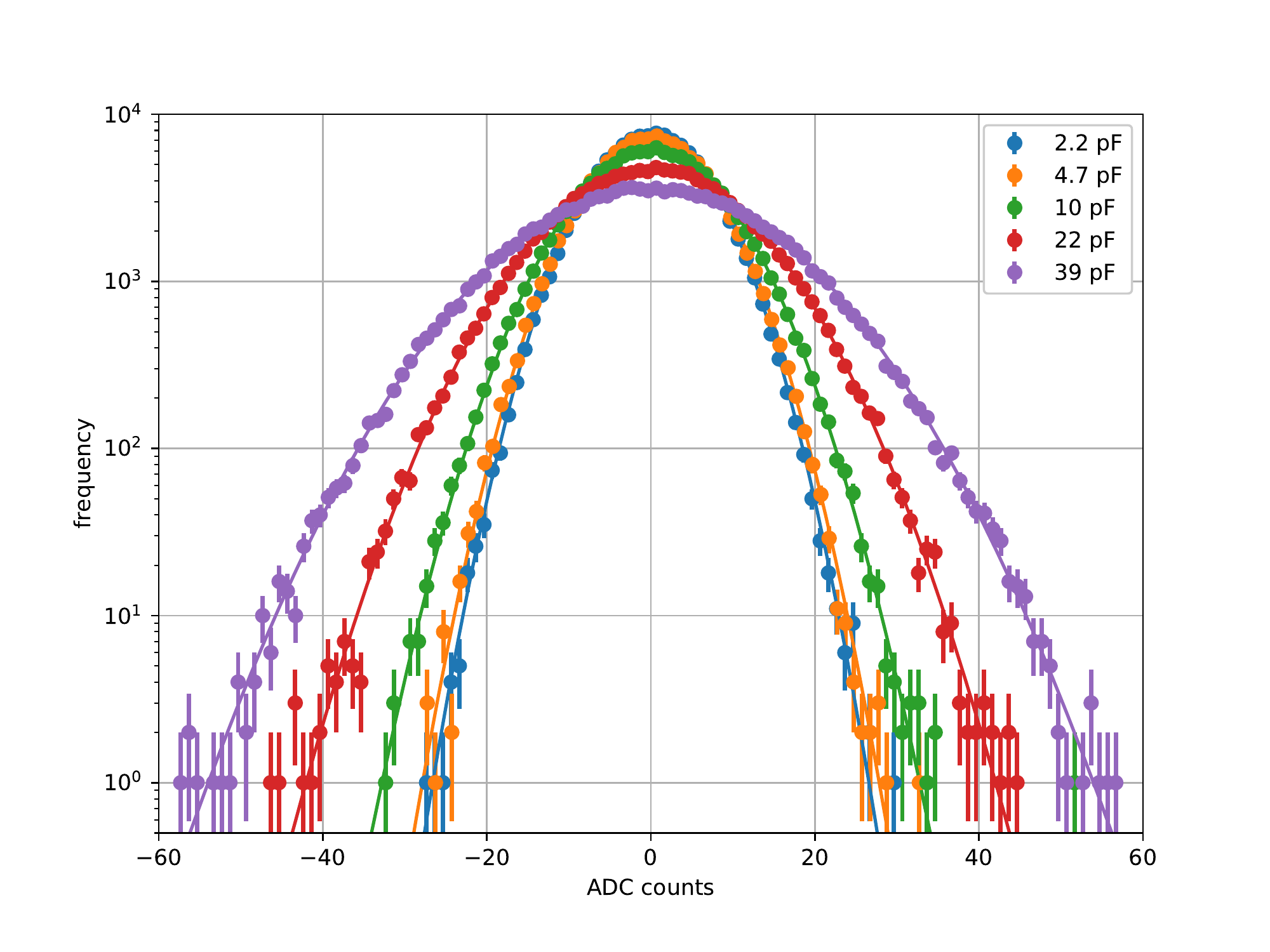}
\includegraphics[width=0.5\textwidth]{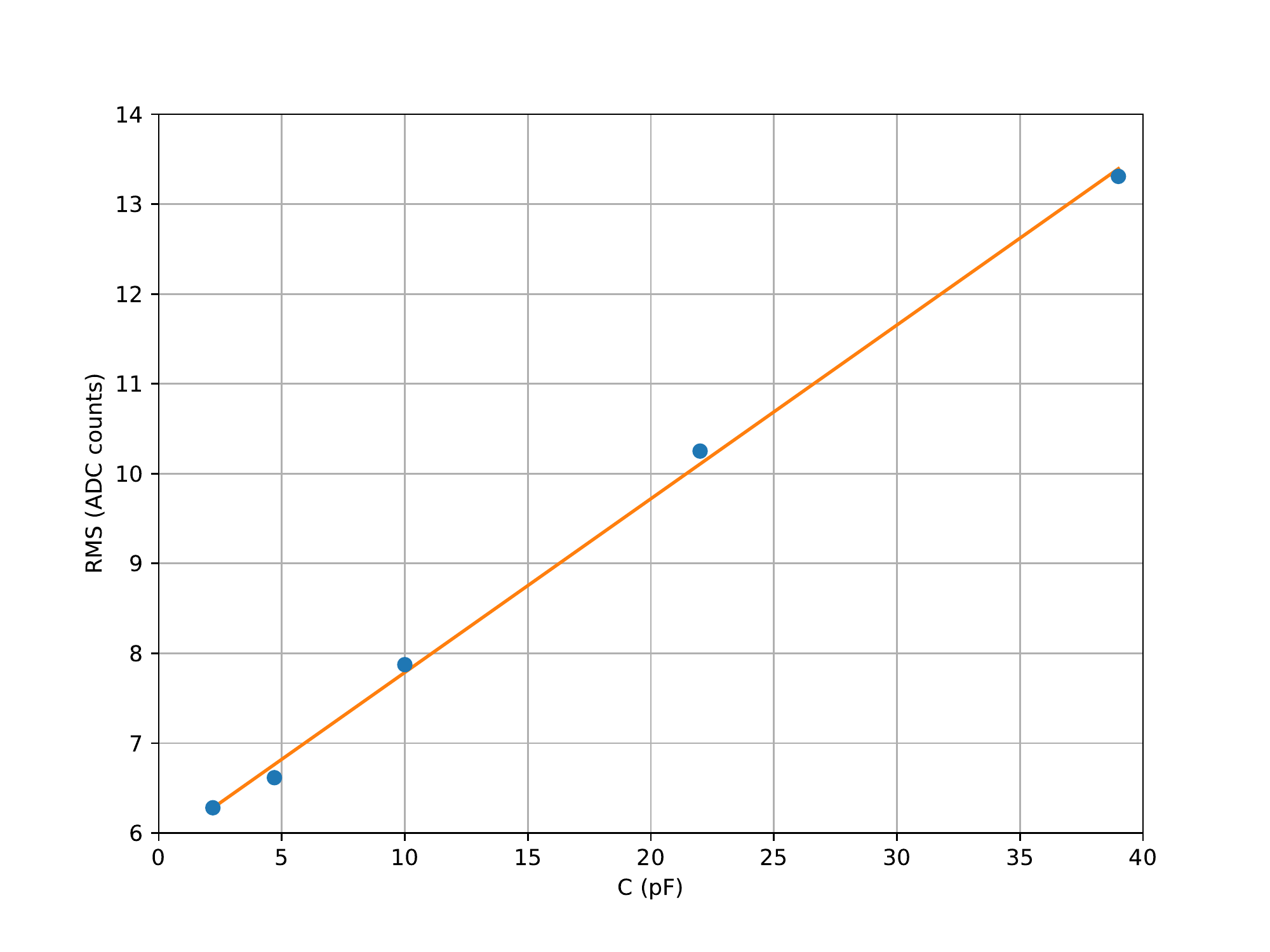}
\caption{\textbf{Top:} Histogram of data taken by externally triggering the ASIC while one of its inputs was under capacitative load. 
The mean ADC value (\ie\ the pedestal) has been subtracted. 
Each distribution is well-defined by a Gaussian, also shown as a colour-coded line.
\textbf{Bottom:} RMS ($\sigma$) of the above histograms as a function of capacitor value. 
The $x$-axis does not include any additional capacitance due to (\eg) the ASIC test board itself, so the $y$-intercept of this line is an upper limit on the noise of the ASIC.
}
\label{fig:noisedist}
\end{figure}

An example noise measurement for one channel is given in the top panel of \cref{fig:noisedist}; each distribution is well-approximated by a Gaussian distribution. 
The RMS of each distribution is plotted as a function of the load capacitance in the bottom panel of \cref{fig:noisedist}. 
As expected, the data are linear with load capacitance. 
Note that in this setup, it is only possible to measure the \emph{slope} of this curve, since the $x$-axis includes only the load capacitance and not any additional capacitance due to (\eg) traces on the ASIC circuit board between the input headers and input wire-bond pads. 

\begin{figure}[bthp]
  \begin{center}
    \includegraphics[width=0.5\textwidth,  trim={1.5cm 2cm 1.5cm 0cm}, clip]{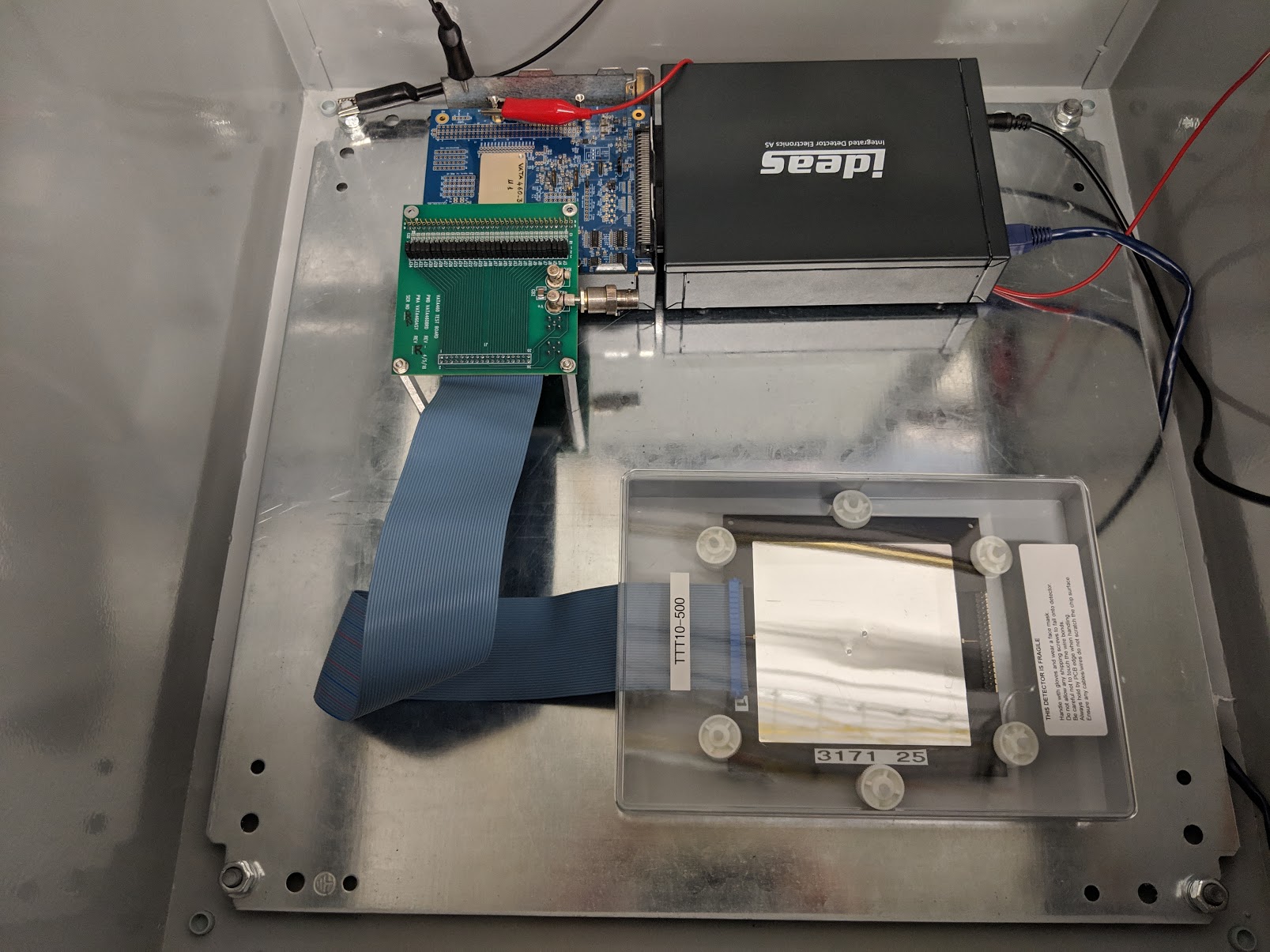}
  \end{center}
  \caption{Image of an IDEAS VATA460.3 development board (top right board) connected to a ``Galao'' readout FPGA (top right with IDEAS logo); the Galao is a universal readout board which can be used with several different IDEAS ASIC test boards which provides an Ethernet interface to the ASIC. 
  It also includes functionality for (\eg) external triggering and charge injection into the ASIC. 
  The ASIC itself is contained beneath the large shield in the middle of the test board. 
  In this setup, the ASIC inputs are connected to an interface board which, in turn, is connected to a single-sided silicon strip detector (bottom) via a ribbon cable, allowing the ASIC to digitize signals from the detector.
   \label{fig:asic}
}
\end{figure}

\begin{figure}[hbtp]
  \begin{center}
    \includegraphics[width=0.5\textwidth, trim={0 0 0 0.95cm}, clip]{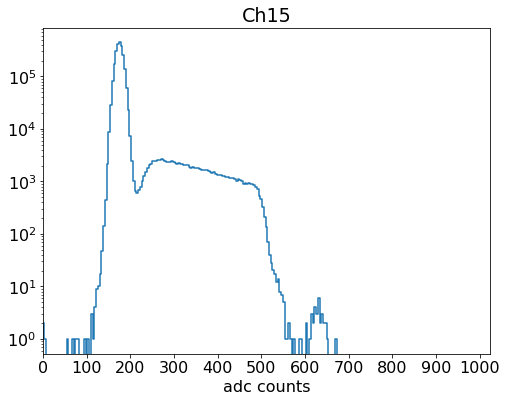}
  \end{center}
  \caption{
  \ce{^{137}Cs} spectrum taken with the IDEAS ASIC and a single-sided silicon detector. 
  The Compton edge (\SI{500}{counts}) and the beginnings of a photo-absorption peak  (\SI{630}{counts}) can be seen. 
  Fitting this peak with a Gaussian distribution yields a width of \SI{28(6)}{counts}, corresponding to a FWHM of $\sim 11\%$.
  This is relatively high due to the parasitic capacitance of this particular setup. 
  }

  \label{fig:Cs137}
\end{figure}

While the DSSDs are being manufactured, the ASIC is being tested to read out single-sided silicon detectors already located at Goddard; these detectors have already been fully characterized so will provide a useful reference moving forward. 
Furthermore, this allows for software and procedures required for testing to be developed prior to the arrival of the \compair\ silicon. 
A photo of the test setup is given in \cref{fig:asic}.
The results of the very first data measured using a \ce{^{137}Cs} source are given in \cref{fig:Cs137}.
Note that these data were taken with detectors very different from the prototype detectors and hence the exact ASIC configuration used here is different from the one that will be used using the DSSDs.
The noise in these measurements is also relatively high ($\sim 11\%$ FWHM for the photo-absorption peak) due to the exact configuration used in this setup; tests of the prototype DSSDs are expected to have better performance due to lower parasitic capacitance in the detector/ASIC interface. 

\section{Conclusions and Timeline}

\amego\ is a powerful \SI{}{\MeV} gamma-ray telescope that will open a window on the universe with unprecedented sensitivity. 
A prototype telescope is under development at NASA Goddard and the US Naval Research Lab. 
Over the next few months, DSSDs will arrive at Goddard and undergo testing; the associated front end electronics for the silicon subsystem are being designed in parallel. 
Development of a fully-custom front-end (including a custom ASIC) is also under investigation in preparation for full \amego.
The other \compair\ subsystems are also being developed; over the next year the subsystems will be integrated and beam tested in preparation for a balloon flight.

\end{document}